\documentclass[11pt,aps,prc,preprintnumbers,amsmath,amssymb,
epsf,superscriptaddress,groupedaddress,nofootinbib,tightenlines]{revtex4}

\usepackage{graphicx}
\usepackage{amsmath}
\usepackage{bm}
\usepackage{bbold}
\usepackage{color}

\def\beqn{\begin{eqnarray}}
\def\eeqn{\end{eqnarray}}
\def\barr{\begin{array}}
\def\earr{\end{array}}
\def\btab{\begin{tabular}}
\def\etab{\end{tabular}}
\def\bite{\begin{itemize}}
\def\eite{\end{itemize}}
\def\bcen{\begin{center}}
\def\ecen{\end{center}}

\def\eq{\begin{equation}}
\def\ee{\end{equation}}
\def\nn{\nonumber}

\def\q2dagger{q_2\hspace{-0.35cm}/\;}

\begin{document}

\preprint{MITP/15-079}
\preprint{NT@UW-15-13}

\title{How strange is pion electroproduction?}
\author{Mikhail Gorchtein}
\affiliation{
PRISMA Cluster of Excellence, 
Institut f\"ur Kernphysik, 
Johannes Gutenberg-Universit\"at, Mainz, Germany
}
\email{gorshtey@kph.uni-mainz.de}

\author{Hubert Spiesberger}
\affiliation{
PRISMA Cluster of Excellence, 
Institut f\"ur Physik, 
Johannes Gutenberg-Universit\"at, Mainz, Germany
}
\affiliation{
Centre for Theoretical and Mathematical Physics 
and Department of Physics, 
University of Cape Town, Rondebosch 7700, South Africa
}
\email{spiesber@uni-mainz.de}

\author{Xilin Zhang}
\affiliation{
Department of Physcs, University of Washington, 
Seattle, Washington, USA
}
\email{xilinz@uw.edu}

\begin{abstract}
We consider pion production in parity-violating electron scattering 
(PVES) in the presence of nucleon strangeness in the framework of 
partial wave analysis with unitarity. Using the experimental 
bounds on the strange form factors obtained in elastic PVES, we 
study the sensitivity of the parity-violating asymmetry to 
strange nucleon form factors. For forward kinematics and electron 
energies above 1 GeV, we observe that this sensitivity 
may reach about 20\% in the threshold region. With parity-violating 
asymmetries being as large as tens p.p.m., this study suggests 
that threshold pion production in PVES can be used as a promising 
way to better constrain strangeness contributions. Using this model 
for the neutral current pion production, we update the estimate 
for the dispersive $\gamma Z$-box correction to the weak charge 
of the proton. In the kinematics of the Qweak experiment, our new 
prediction reads Re$\,\Box_{\gamma Z}^V(E=1.165\,{\rm GeV}) = 
(5.58\pm1.41)\times10^{-3}$, an improvement over the previous 
uncertainty estimate of $\pm2.0\times10^{-3}$. Our new prediction 
in the kinematics of the upcoming MESA/P2 experiment reads 
Re$\,\Box_{\gamma Z}^V(E=0.155\,{\rm GeV}) = (1.1\pm0.2) \times 
10^{-3}$.
\end{abstract}
\date{\today}

\maketitle


\section{Introduction}

The discovery of weak neutral current interactions in 
parity-violating electron scattering (PVES) \cite{Prescott:1978tm, 
Prescott:1979dh} and in atomic parity violation (APV) 
\cite{Wood:1997zq} provided an important proof for the structure 
of the Standard Model (SM). The accuracy of modern experiments 
provides access to physics beyond the Standard Model (BSM) in 
a mass range which is comparable or complementary to searches 
at colliders and in astrophysics \cite{Erler:2004in,Erler:2013xha}. 
Two experiments designed to test the SM running of the weak 
mixing angle at low energies, Qweak at the Jefferson Laboratory 
in the U.S. \cite{Androic:2013rhu} and P2@MESA at Mainz University, 
Germany \cite{Becker:2014hta} will constrain SM extensions with 
mass scales in the 30 - 50 TeV range. 

The interpretation of these high-precision experiments in terms 
of the fundamental SM parameters is based on a similarly precise 
calculation of electroweak radiative corrections 
\cite{Marciano:1983ss,Erler:2004in}. At the one-loop level, 
$\gamma Z$-box graph corrections constitute a numerically 
important contribution. Their evaluation requires knowledge of 
the hadron structure at low energies, i.e.\ in a region where 
perturbation theory can not be applied. Recently, the vector 
part of these corrections was re-evaluated in the framework 
of forward dispersion relations \cite{Gorchtein:2008px}, and its 
value and uncertainty was found to be considerably larger than 
previously anticipated. Subsequent work allowed to constrain 
its central value \cite{Gorchtein:2008px, Sibirtsev:2010zg, 
Rislow:2010vi,Gorchtein:2011mz,Hall:2013hta}, but the size of 
its uncertainty is still an open question. For the kinematics 
of the Qweak and P2 experiments, the dispersion representation 
of the $\gamma Z$-box graph correction involves the inclusive 
inelastic interference structure functions $F_{1,2}^{\gamma Z}$ 
integrated over the full kinematical range with a strong 
emphasis on the low-energy range, $Q^2 \leq 1$ GeV$^2$ and 
$W \leq 4$ GeV ($Q^2$ is the virtuality of the space-like 
photon or $Z$-boson originating from electron scattering, 
and $W$ the invariant mass of the hadronic final state $X$ 
resulting from the process $\gamma^* + N \to X$ with $X = 
\pi N,\,2\pi N$, etc.)\footnote{
The situation is different for the axial-vector part of 
$\gamma Z$-box graph corrections which involve $F_{3}^{\gamma Z}$. 
In this case, loop momenta $\sim M_Z$ are dominating and the 
calculation can be performed in a reliable way within perturbation 
theory \cite{Marciano:1983ss}. 
}. 

The approach of Refs.\ \cite{Gorchtein:2008px,Sibirtsev:2010zg, 
Rislow:2010vi,Gorchtein:2011mz,Hall:2013hta}, in the absence of 
any detailed interference data in the required kinematical 
range, was based on a purely phenomenological fit to the 
electromagnetic inelastic total cross section data 
\cite{Christy:2007ve}, complemented with an isospin-rotation 
to predict $F_{1,2}^{\gamma Z}$. Unfortunately, this procedure 
is essentially {\it ad hoc}, and leads to model-dependent 
uncertainty estimates, reflected by the spread of the 
uncertainties given in Refs.\ \cite{Rislow:2010vi,
Gorchtein:2011mz,Hall:2013hta}. 

In the present work we try to construct the input to the 
dispersion relations, starting at threshold for pion production, 
in a more controlled way. In this range, approximately determined 
by $M + m_\pi \leq W\leq2$ GeV and $Q^2 \leq 2$ GeV$^2$, one can 
rely on very detailed experimental data for pion photo- and 
electroproduction that allow for partial wave analyses as 
implemented in MAID \cite{Drechsel:1998hk, Drechsel:2007if} and 
SAID \cite{SAID}. In this approach it is also possible to 
explicitly take account of constraints due to unitarity and 
symmetries and include dynamical effects of strong rescattering.  

In the literature, the weak pion production amplitudes have been 
constructed from the electromagnetic ones, and observables have 
been studied upon neglecting strangeness contributions 
\cite{Mukhopadhyay:1998mn,Sato:2003rq}. The main distinction of 
the present work is in avoiding this assumption. This leads to a 
natural uncertainty estimate due to strangeness contributions 
which, firstly, is driven by experimental data on strange form 
factors from elastic PVES \cite{Armstrong:2012bi}, and, secondly, 
brings this uncertainty estimate in direct correspondence with 
that of inelastic PVES data. Such data above the pion production 
threshold and reaching into the $\Delta$-resonance region have 
been taken by the G0 Collaboration at JLab \cite{Androic:2012doa} 
and by the A4 Collaboration at MAMI, Mainz \cite{A4resPV}. Our 
formalism can also serve as a basis for extracting the strange 
form factors from threshold pion production in PVES experiments. 
The advantage of this method lies in the fact that PV asymmetries 
are large, in the range of several tens of p.p.m. as opposed to 
a few p.p.m. for asymmetries in elastic PVES, which were 
traditionally used to access strangeness contributions. 

Another closely related topic concerns hadronic parity violation 
that leads to induced PV contributions in electromagnetic 
interactions and may be seen in PVES as well as in parity 
violation in nuclei \cite{Zhu:2000gn,Haxton:2013aca}. In elastic 
PVES, these contributions manifest themselves in a similar way 
as effects from the axial-vector coupling of the $Z$ boson at 
the hadronic side, but can be disentangled from 
the $Z$-exchange contribution in PV pion electroproduction due to a different $Q^2$-dependence.  Ref. \cite{Chen:2000km} used the "DDH best value" of the PV $\pi NN$ coupling constant $h^1_\pi$ \cite{Desplanques:1979hn}, and showed that PV threshold $\pi^+$ electroproduction at low energies and forward angles is very sensitive to this coupling. On the other hand, there are indications that the actual value of $h^1_\pi$ is at least four times smaller \cite{Page:1987ak}. Having in mind such contributions, we will focus on electron energy range $\sim1$ GeV and not too forward angles. We postpone the detailed discussion of the interplay of hadronic PV effects with the strangeness to the upcoming work. 

The article is organized as follows. In Section 
\ref{sec:kinematics} we lay out the formalism and define the 
kinematics, in Section \ref{sec:multipoles} we explicitly 
construct the multipoles with the weak vector current and 
incorporate strangeness contributions. Section \ref{sec:apv} 
deals with the sensitivity of the PV asymmetry in inelastic 
PVES to strange form factors; in Section \ref{sec:gammaZ} we 
apply the model for weak pion production developed in the 
previous sections to the calculation of the dispersion 
$\gamma Z$-correction to the proton's weak charge. 
Section \ref{sec:conclusions} contains our concluding remarks.


\section{Kinematics and definitions}
\label{sec:kinematics}
\begin{figure}[t]
\includegraphics[width=6cm]{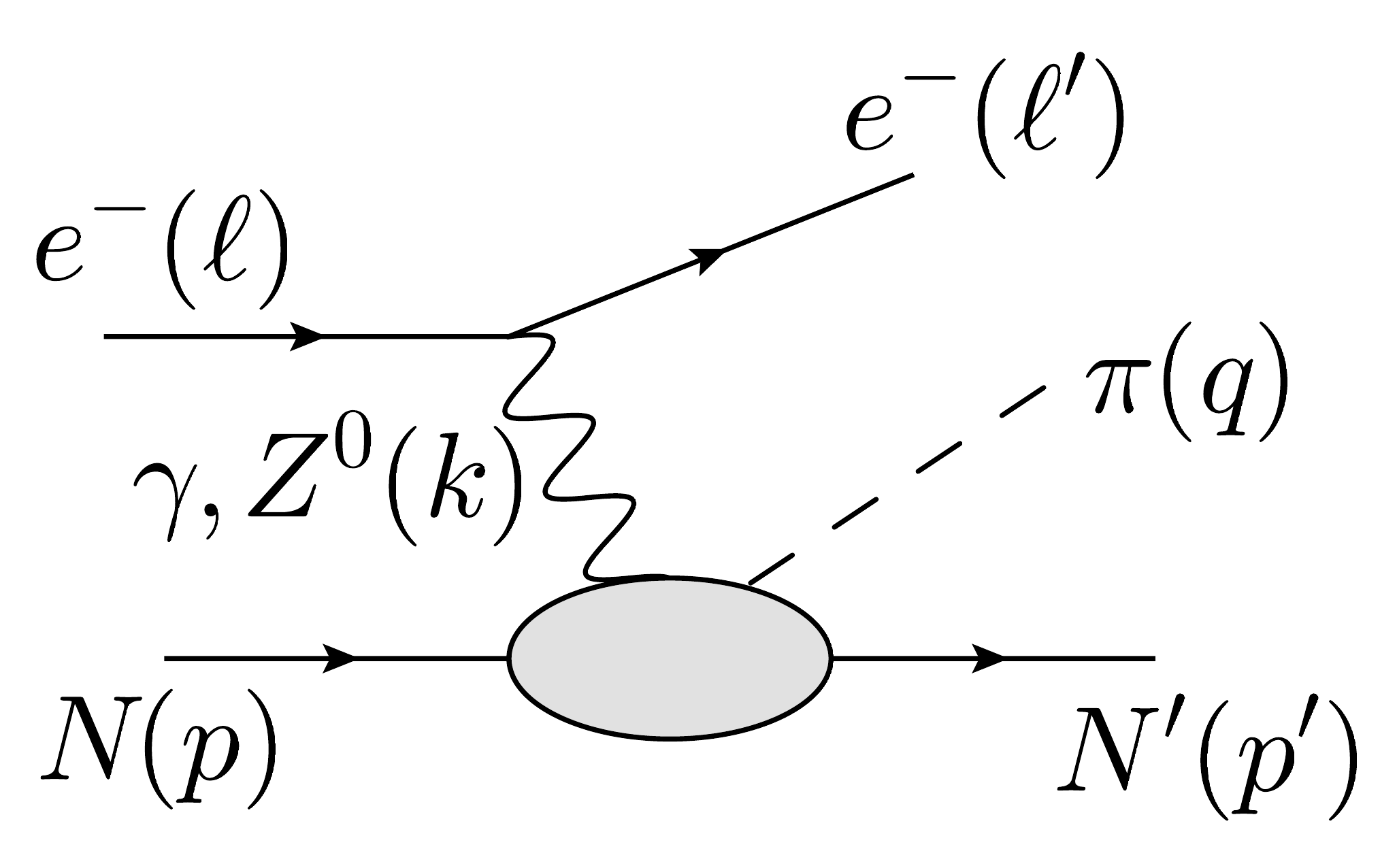}
\caption{Pion electroproduction via a neutral-current interaction 
described by the exchange of a virtual photon or a $Z$ boson.}
\label{fig:diagram}
\end{figure}

In this work we consider pion electroproduction off a nucleon 
of mass $M$, $e^-(\ell) + N(p) \to e^-(\ell') + \pi(q)+N'(p')$, 
as shown in Fig.\  \ref{fig:diagram}. The interaction is 
described by diagrams with the exchange of one boson which 
carries the four-momentum $k=\ell-\ell'$, 
\beqn
T &=& 
\frac{e^2}{Q^2} 
\bar u(\ell')\gamma^\mu u(\ell) 
\langle N,\pi|J_\mu^{EM}|N\rangle
\eeqn
for the contribution of the electromagnetic interaction, and 
\beqn
T &=& 
- \frac{G_F}{2\sqrt2} 
\bar u(\ell')\gamma^\mu(g_V^e+g_A^e\gamma_5) u(\ell) 
\langle N,\pi|J_\mu^{NC}-J_{\mu5}^{NC}|N\rangle
\label{eq:NC}
\eeqn
for the weak neutral current (NC) interaction. We have defined 
$Q^2 = -k_\mu k^\mu>0$. The kinematics of the reaction 
$\left( \gamma/ Z \right)(k) + N(p) \to \pi(q) + N'(p')$ in the 
$\pi N$ center of mass frame is completely fixed in terms of 
three Lorentz scalars for which we take the invariant mass of 
the hadronic final state $W$, $W^2 = (p'+q)^2 = (p+k)^2$, the 
virtuality $Q^2$ of the initial boson, and the four-momentum 
transfer to the nucleon $t = (p'-p)^2 < 0$. In the following, 
we will use the center-of-mass frame of the initial 
nucleon-photon pair (or, equivalently, of the final 
nucleon-pion pair) defined by $\vec k + \vec p = \vec q + 
\vec{p}^{\,\prime} = 0$. In this reference frame, the 
kinematics can be specified by the energy of the photon, 
$\omega$, its virtuality $Q^2$, and the pion scattering angle 
$\theta$. We parametrize the momentum 4-vectors by 
\beqn
k^\mu &=& (k^0,0,0,k) \, , 
\quad 
p^\mu = (E_1,0,0,-k) \, , 
\nn \\
q^\mu &=& (E_\pi,q\sin\theta,0,q\cos\theta) \, , 
\quad 
p^{\, \prime\mu} = (E_2,-\vec q\,) \, . 
\eeqn
with 
\beqn
k^0 &=& 
\frac{W^2-M^2-Q^2}{2W} \, , 
\quad 
E_1 = \frac{W^2+M^2+Q^2}{2W} \, , 
\nn \\
E_\pi &=& \frac{W^2-M^2+m_\pi^2}{2W} \, , 
\quad 
E_2 = \frac{W^2+M^2-m_\pi^2}{2W} \, , 
\nn \\
|\vec k| &\equiv & k = 
\frac{\sqrt{[(W-M)^2+Q^2][(W+M)^2+Q^2]}}{2W} \, , 
\nn \\
|\vec q| &\equiv & q = 
\frac{\sqrt{[W^2-(M+m_\pi)^2][W^2-(M-m_\pi)^2]}}{2W} \, , 
\nn \\
\cos\theta &=& \frac{2k^0E_\pi+t+Q^2-m_\pi^2}{2kq} \, .
\eeqn


\subsection{Invariant amplitudes}

The invariant amplitudes with the vector current were introduced, 
e.g., in Ref.\ \cite{Adler:1968tw} as 
\beqn
\langle N,\pi|J_\mu^{EM,\,NC}|N\rangle 
&=& 
\sum_{i=1}^6V^{\gamma,Z}_i(W^2, Q^2, t) 
\bar U_f O^\mu_{V_i}U_i \, ,
\eeqn
with
\beqn
O^\mu_{V_1} &=& 
\frac{i}{2}\left(\gamma^\mu\not \!k\,-\not \!k\,\gamma^\mu\right)\gamma_5 \, , 
\nn\\
O^\mu_{V_2} &=& 
-2i\left(P^\mu(q\cdot k)-q^\mu(P\cdot k)\right)\gamma_5 \, , 
\nn\\
O^\mu_{V_3} &=& 
\left(\gamma^\mu(q\cdot k)-q^\mu\not\!k\right)\gamma_5 \, , 
\nn\\
O^\mu_{V_4} &=& 
2\left(\gamma^\mu(P\cdot k)-P^\mu\not\!k\right)\gamma_5 
- 2MO^\mu_{V_1} \, , 
\nn\\
O^\mu_{V_5} &=& 
-i\left(k^\mu(q\cdot k)-q^\mu k^2\right)\gamma_5 \, , 
\nn\\
O^\mu_{V_6} &=& 
\left(k^\mu\not\!k-\gamma^\mu k^2\right)\gamma_5 \, ,
\eeqn
where $U_i\,(U_f)$ stand for the initial (final) nucleon Dirac spinor, and we have used the average nucleon four-momentum $P=(p+p')/2$. 
Similarly, the axial-vector part is separated as 
\beqn
\langle N,\pi|J_{\mu5}^{NC}|N\rangle 
&=& 
\sum_{i=1}^8A^{Z}_i(W^2, Q^2, t) 
\bar U_f O^\mu_{A_i}U_i \, , 
\eeqn
with
\beqn
O^\mu_{A_1} &=& 
-\frac{i}{2}\left(\gamma^\mu\not \!q\,-\not \!q\,\gamma^\mu\right) \, , 
\nn\\
O^\mu_{A_2} &=& 
2iP^\mu \, , 
\nn\\
O^\mu_{A_3} &=& 
iq^\mu \, , 
\nn\\
O^\mu_{A_4} &=& 
-M\gamma^\mu \, , 
\nn\\
O^\mu_{A_5} &=& 
2P^\mu \not\!k \, , 
\nn\\
O^\mu_{A_6} &=& 
q^\mu \not\!k \, , 
\nn\\
O^\mu_{A_7} &=& 
ik^\mu \, , 
\nn\\
O^\mu_{A_8} &=& 
k^\mu\not\!k \, .
\eeqn
The scalar amplitudes $V_i^{\gamma, Z}$, $A_i^Z$ are functions of the invariants $W^2$, $Q^2$ and $t$. The last two structures $O^\mu_{A_{7,8}}$ are lepton mass terms and do not contribute to the neutral current process studied here.


\subsection{Multipole decomposition}

It is common to evaluate the covariant tensors introduced in 
the previous subsection in the center-of-mass frame of the pion 
and the final nucleon, relating the invariant amplitudes to the 
CGLN amplitudes \cite{CGLN} 
\beqn
\epsilon_\mu \sum_{j=1}^6V^{\gamma,Z}_j(W^2, Q^2, t) \bar U_f O^\mu_{V_j}U_i
&=&\sum_{j=1}^6{\cal{F}}_j\chi^\dagger_f\Sigma_j\chi_i,\nn\\
\epsilon_\mu \sum_{j=1}^6A^{Z}_j(W^2, Q^2, t) \bar U_f O^\mu_{A_j}U_i
&=&\sum_{j=1}^6{\cal{G}}_j\chi^\dagger_f\tilde\Sigma_j\chi_i,
\eeqn
with $\epsilon_\mu$ the photon or $Z$ polarization vector, ${\cal{F}}_j,{\cal{G}}_j$ scalar amplitudes, $\chi_i(\chi_f)$ the Weyl spinors for initial (final) nucleon, respectively, and $\Sigma_j,\tilde\Sigma_j$ matrices in the spinor space. The CGLN amplitudes allow for a decomposition into multipoles, 
\beqn
{\cal{F}}_j&=&\sum_l \left\{ a_{j,l\pm}E_{l\pm}+b_{j,l\pm}M_{l\pm}+c_{j,l\pm}S_{l\pm}\right\},\nn\\
{\cal{G}}_j&=&\sum_l \left\{ \tilde a_{j,l\pm}{\cal{E}}_{l\pm}+\tilde b_{j,l\pm}{\cal{M}}_{l\pm}+\tilde c_{j,l\pm}{\cal{S}}_{l\pm}\right\}.\eeqn
Above, $E_{l\pm},M_{l\pm},S_{l\pm}$ are the multipoles describing the vector electric, magnetic and scalar transition to the $\pi N$-state with the angular orbital momentum $l$ and the total orbital momentum $j=l\pm1/2$, and similarly ${\cal{E}}_{l\pm},{\cal{M}}_{l\pm},{\cal{S}}_{l\pm}$ describe transitions with the axial vector interaction. The multipoles are functions of $W$ and $Q^2$ only, and the angular dependence in terms of Legendre polynomials and their derivatives is contained in the coefficients $a,b,c,\tilde a, \tilde b, \tilde c$. The explicit form of the coefficients of the matrices $a,b,c$ for the vector case can be found in Ref.\ \cite{Berends:1967vi} and $\tilde a, \tilde b, \tilde c$ in Ref. \cite{Adler:1968tw} for the axial-vector case. 


\subsection{Isospin structure}

Isospin is not conserved by the electromagnetic interaction. The 
one-photon exchange diagram has, therefore, an isoscalar and an 
isovector component. Combining $I = 0, 1$ of the photon and $I = 
1$ of the pion would lead to three possible isospin amplitudes 
for $I = 0, 1, 2$. From the $t$-channel perspective, i.e.\ for 
pion photoproduction $\gamma\pi\to N\bar N$, however, only $I = 
0$ or $I = 1$ are possible. As a result, there are three 
independent isospin channels which we denote by $M^0$ and 
$M^\pm$. The scattering amplitude can thus be separated as
\beqn
M &=& 
M^0 \bar\chi_f(\vec\tau\cdot\vec\pi)\chi_i 
+ M^+ \bar\chi_f\frac{1}{2}\{(\vec\tau\cdot\vec\pi),\tau_3\}\chi_i 
+ M^- \bar\chi_f\frac{1}{2}[(\vec\tau\cdot\vec\pi),\tau_3]\chi_i 
\, ,
\eeqn
with $\chi_{i,f}$ the two-component nucleon's isospinors, 
$\vec \tau$ the Pauli matrices and $\vec \pi$ denote the 
pion isovectors. In terms of  these isospin-channel amplitudes, 
charge-channel amplitudes are expressed as 
\beqn
M^{\pi^+n} &=& \sqrt2(M^0+M^-) \, ,
\nn\\
M^{\pi^-p} &=& \sqrt2(M^0-M^-) \, ,  
\nn\\
M^{\pi^0p} &=& M^+ + M^0 \, , 
\nn\\
M^{\pi^0n} &=& M^+ - M^0 \, . 
\label{eq:iso_gapi}
\eeqn
These relations are equally valid for multipole, invariant or 
CGLN amplitudes. 

The isospin decomposition of the electromagnetic and weak neutral 
current in terms of quark currents (we consider the three 
lightest flavors $u,d,s$ only) reads 
\beqn
J_\mu^{EM} &=& 
\sum_{q} e^q \bar u_q\gamma_\mu u_q = 
\bar q \gamma_\mu 
\left(\frac{1}{6}\mathbb{1} + \frac{1}{2}\tau_3 \right)q 
- \frac{1}{3}\bar s\gamma_\mu s \, , 
\nn\\
J_\mu^{NC} &=& 
\sum_{q\in N} g_V^q \bar u_q\gamma_\mu u_q = 
\bar q \gamma_\mu 
\left(\xi_V^{I=0}\frac{1}{6}\mathbb{1} 
+ \xi_V^{I=1}\frac{1}{2}\tau_3 \right)q + 
\xi_V^{s}\bar s\gamma_\mu s \, , 
\nn\\
J_{\mu5}^{NC} &=& 
\sum_{q\in N} g_A^q \bar u_q\gamma_\mu\gamma_5 u_q = 
\bar q \gamma_\mu\gamma_5 
\left(\xi_A^{I=0}\frac{1}{6}\mathbb{1} + 
\xi_A^{I=1}\frac{1}{2}\tau_3 \right)q 
+ \xi_A^{s}\bar s\gamma_\mu\gamma_5 s \, ,
\eeqn
with $q = \left(\begin{array}{c}u\\d\end{array}\right)$. At tree 
level in the Standard Model and according to the normalization 
defined in Eq.\ (\ref{eq:NC}), the coupling constants appearing 
in these equations are determined by the weak mixing angle, 
$\sin^2\theta_W$, and given by $\xi_V^{I=1} = 2 - 4\sin^2\theta_W$, 
$\xi_V^{I=0} = - 4\sin^2\theta_W$, $\xi_A^{I=1} = 2$, 
$\xi_A^{I=0} = 0$, $\xi_V^s = -1/2 + (2/3)\sin^2\theta_W$, 
and $\xi_A^s = -1/2$. From this decomposition we obtain the 
standard expressions for the weak form factors of the nucleon,
\beqn
G_{E,M}^{Z,\,p} &=& 
- G_{E,M}^{\gamma,\,n} 
+ (1-4\sin^2\theta_W)G_{E,M}^{\gamma,\,p} 
- G_{E,M}^{s} \, , 
\nn\\
G_{E,M}^{Z,\,n} &=& 
- G_{E,M}^{\gamma,\,p} 
+ (1-4\sin^2\theta_W)G_{E,M}^{\gamma,\,n} 
- G_{E,M}^{s} \, . 
\label{eq:iso_Z}
\eeqn
In the same way, the flavor decomposition of the amplitudes 
for weak vector pion production, identifying, e.g., $M^- = 
\frac{1}{\sqrt2}(M^{\pi^+n} - M^{\pi^-p})$, and keeping 
strangeness, can be decomposed as 
\beqn
M^{\pi^+n}_Z &=& 
- M^{\pi^-p}_\gamma 
+ (1-4\sin^2\theta_W)M^{\pi^+n}_\gamma - \sqrt2M_s \, , 
\nn\\
M^{\pi^-p}_Z &=& 
- M^{\pi^+n}_\gamma 
+ (1-4\sin^2\theta_W)M^{\pi^-p}_\gamma - \sqrt2M_s \, , 
\nn\\
M^{\pi^0p}_Z &=& 
M^{\pi^0n}_\gamma 
+ (1-4\sin^2\theta_W)M^{\pi^0p}_\gamma - M_s \, , 
\nn\\
M^{\pi^0n}_Z &=& 
M^{\pi^0p}_\gamma 
+ (1-4\sin^2\theta_W)M^{\pi^0n}_\gamma + M_s \, . 
\label{eq:iso_Zpi}
\eeqn
The presence of the strangeness contribution is the main distinction 
of this analysis from other calculations of pion production in 
electroweak reactions.


\section{Vector multipoles from isospin symmetry and strangeness 
form factors}
\label{sec:multipoles}

Upon neglecting the strangeness contribution, the vector multipoles 
in weak NC pion-production is given exactly in terms of the 
electromagnetic multipoles given in Eq.\ (\ref{eq:iso_Zpi}). 
To our knowledge, this is how pion production is dealt with 
in all phenomenological models that relate pion production in 
electromagnetic and weak NC reactions. The aim of this section 
is to go beyond this approximation and model the strangeness 
contributions. 

\begin{figure}[t]
\includegraphics[width=16cm]{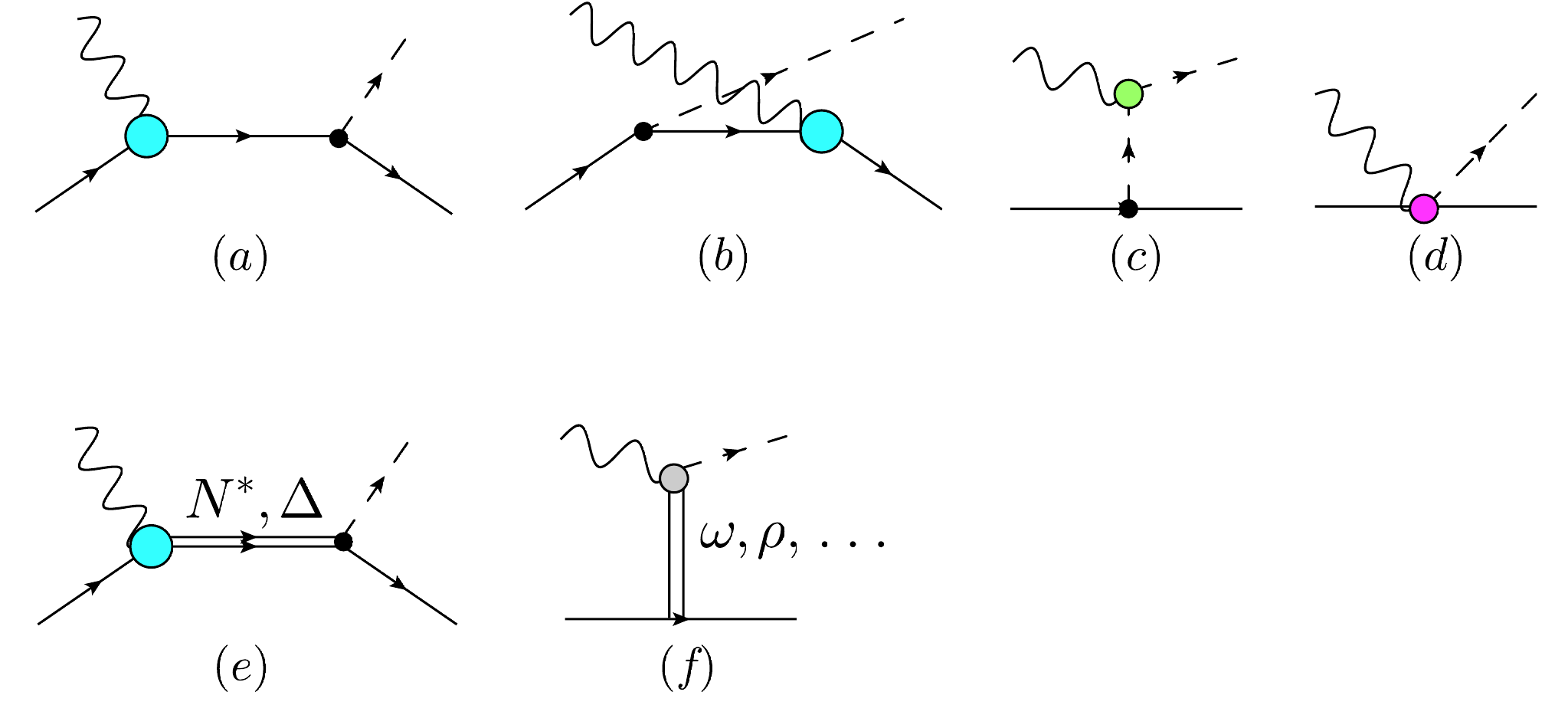}
\caption{
Isobar model for pion-production. (a) -(d) are Born contributions, 
(e) are resonance contributions and the $t$-channel vector 
exchange diagrams are shown in (f). The blobs indicate the elastic 
nucleon and pion, and various transition form factors.
}
\label{fig:Born}
\end{figure}

We start from the isobar model as implemented in MAID 
\cite{Drechsel:1998hk,Drechsel:2007if}. The pion-production 
amplitude can be represented as a sum of tree-level amplitudes, 
as shown in Fig.\ \ref{fig:Born}. These can be modeled in terms 
of tree-level coupling constants and form factors. Rescattering 
effects are incorporated following the unitarization procedure 
in the $K$-matrix approach used in MAID \cite{Drechsel:1998hk}. 
There, the unitarized Born amplitude $M^{B,\alpha}_{\gamma\pi}$ 
is obtained from the tree-level Born amplitude 
$B^{\alpha}_{\gamma\pi}$ and the elastic pion-nucleon scattering 
amplitude $t^{\alpha}_{\pi N} = [\eta_\alpha 
\exp(2i\delta_\alpha) - 1]/2i$ (taken from the SAID 
analysis \cite{Arndt:1995ak}) expressed in terms of phase shifts 
$\delta_\alpha$ and inelasticities $\eta_\alpha$ as 
\beqn
M^{B,\alpha}_{\gamma\pi} 
= 
B^{\alpha}_{\gamma\pi}[1+it^{\alpha}_{\pi N}] \, ,
\eeqn
where the index $\alpha = (j$, $l$, $I$,\dots) carries all relevant 
quantum numbers: total and orbital angular momentum, isospin, 
multipolarity etc. The unitarized Born amplitude defined in 
this way has the phase of the pion-nucleon scattering amplitude 
below the two-pion production threshold and obeys the Watson 
theorem. Note that the Born amplitude described by the sum of the 
graphs (a) - (d) in Fig.\ \ref{fig:Born} is gauge invariant, which 
is achieved by requiring a certain form of the form factor of 
the contact term, graph (d) in this figure  
\cite{Berends:1967vi,Adler:1968tw}. Gauge invariance makes such 
a unitarization procedure for the Born part meaningful. The MAID 
approach consists then in adding further contributions like vector 
meson exchange contributions (graph (f)) within the same 
unitarization procedure, and resonances (graph (e)) with an 
appropriate phase, such that the full amplitude 
$M^{\alpha}_{\gamma\pi} = M^{B,\alpha}_{\gamma\pi} + 
M^{V,\alpha}_{\gamma\pi} + M^{R,\alpha}_{\gamma\pi}$ has the 
correct phase \cite{Drechsel:1998hk}. 

Since the pion is an isovector, only the isovector components of 
the photon and $Z$ boson couple to the pion field. 
The same is true for all $N\to\Delta$ transitions. 
Correspondingly, only $Z NN$ and $Z NN^*$ couplings can obtain 
additional contributions from strangeness. This statement operates 
with "bare" couplings of the photon. However, rescattering 
effects that could modify the tree level amplitudes are due to 
the strong interaction which conserves isospin and flavor. 

The most straightforward way to estimate the strangeness 
contribution is to use the available experimental information 
on strange form factors of the nucleon. A recent global 
analysis \cite{Armstrong:2012bi} gives the following 
values of the strange electric and magnetic form factors of the 
nucleon at $Q^2=0.1$ GeV$^2$,
\beqn
G_M^s(0.1\,{\rm GeV}^2) &=& 0.29\pm0.21 \, , 
\nn\\
G_E^s(0.1\,{\rm GeV}^2) &=& -0.008\pm0.016 \, . 
\label{eq:strangeFF}
\eeqn
For this analysis, the strange form factors were parametrized 
as $G_E^s = \rho_s\tau G_D(Q^2)$ and $G_M^s = \mu_s G_D(Q^2)$, 
with $\tau = Q^2/(4M^2)$, $G_D(Q^2) = (1+Q^2/\Lambda_V^2)^{-2}$, 
and $\Lambda_V=0.84$ GeV. With this parametrization, 
extrapolating $G_M^s$ from $Q^2 = 0.1$ GeV$^2$ to the origin 
one obtains $\mu_s=0.38\pm0.27$. Analyses based on lattice QCD 
tend to give smaller values of the strange form factors with a much 
smaller uncertainty \cite{Shanahan:2014tja,Green:2015wqa}. For 
our purpose, the values given in Eq.\ (\ref{eq:strangeFF}) are 
sufficient, since one of the goals of this work is to propose a 
new way of extracting the strangeness contribution from the 
experimental data. The estimates from lattice QCD are 
automatically included in the explored parameter range. 
We also note that the values given above, if rewritten in terms 
of Dirac and Pauli strange form factors using $G_M^s = F_1^s + 
F_2^s$ and $G_E^s = F_1^s - \tau F_2^s$, are consistent with 
$F_1^s = 0$ and $F_2 = \mu_sG_D(Q^2)$. 

The strange magnetic moment contribution to the Born multipoles 
can now be calculated in a straightforward way, and we make use 
of the expressions for the Born multipoles published in Berends 
et al.\ \cite{Berends:1967vi}. However, we have to extend the 
analysis of \cite{Berends:1967vi} where the $\pi NN$ coupling 
is assumed to be purely pseudoscalar, whereas MAID uses a 
combination of pseudoscalar and pseudovector couplings,
\beqn
\langle N|\pi^a|N\rangle = 
\frac{g_{\pi NN}}{2M} 
\bar N(p')\not \!q\gamma_5\tau^a N(p) 
\frac{\Lambda_m^2}{\Lambda_m^2+{\vec q\,}^2} 
+ g_{\pi NN} \bar N(p')\gamma_5\tau^a N(p) 
\frac{\vec q\,^2}{\Lambda_m^2+\vec q\,^2} \, ,
\label{eq:piNNmaid}
\eeqn
with $\vec q$ the three-momentum of the pion. The parameter 
$\Lambda_m=450$ MeV describes the transition from a pseudo-vector 
coupling at the pion production threshold (where $|\vec q|=0$) 
to a pseudo-scalar coupling at high energies. 
It is this form that we adopt here. As a consequence, an  
additional contact term appears which contributes to four of 
the multipoles, nameley to $E_{0+}$, $M_{1-}$, $S_{0+}$, and 
$S_{1-}$. We present the full expressions for the Born strangeness 
contributions to the multipoles in Appendix A. 

In Figs.\ \ref{fig1} - \ref{fig6} we display results for the 
first few multipoles according to the procedure described above. 
In all figures, the dotted curves represent the MAID results for 
electromagnetic pion production, while the dashed curves include 
the weak neutral current contributions assuming exact isospin 
symmetry with no strangeness, and the solid curves correspond 
to the weak neutral current multipoles including strangeness 
using the strange magnetic moment $\mu_s = 0.38$. Thus the 
difference between the solid and the dashed curves is a measure 
of the sensitivity of the multipoles to the strange magnetic 
moment.
\begin{figure}
\includegraphics[width=16cm]{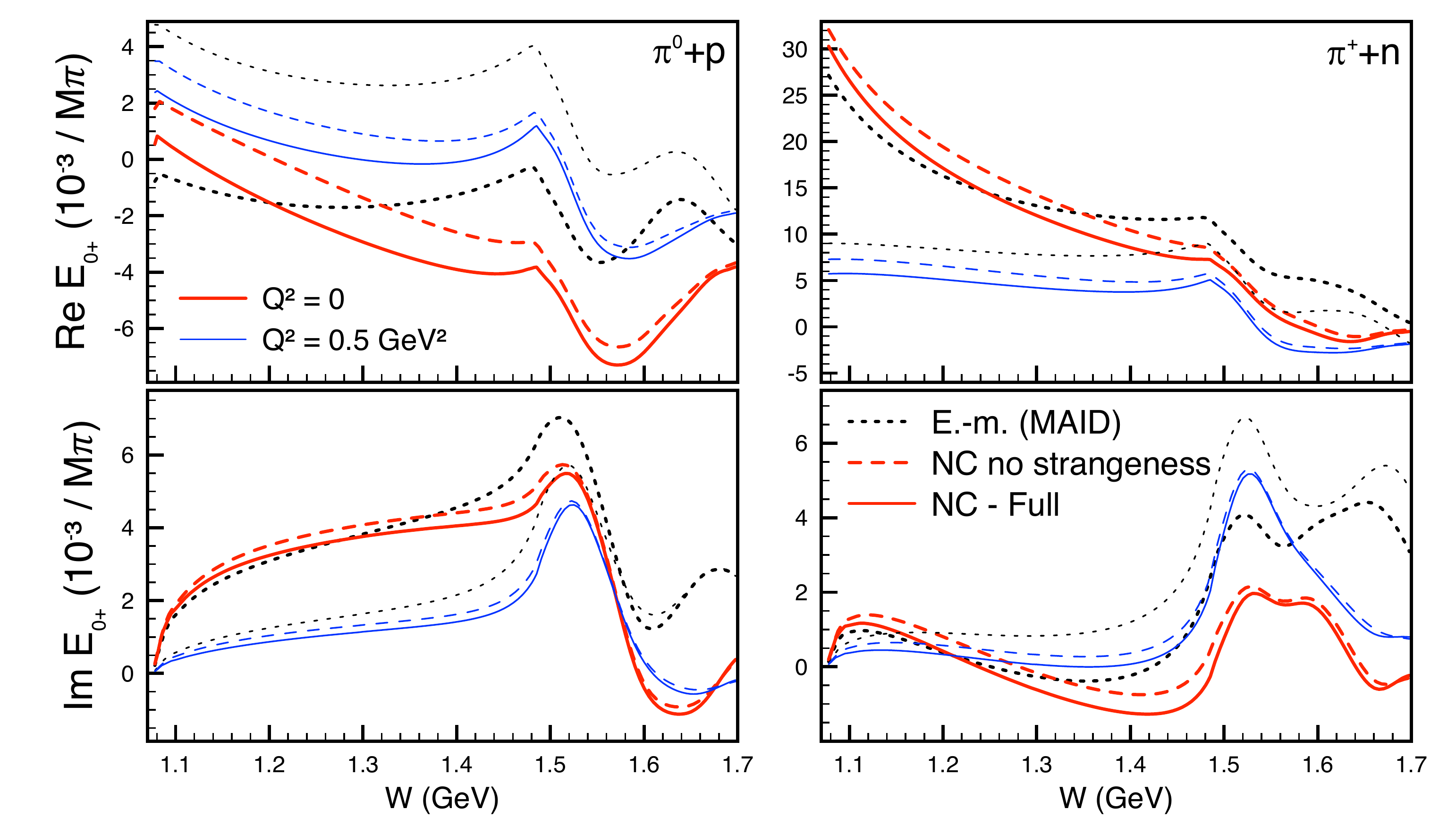}
\caption{
Results for the electromagnetic and weak NC multipole $E_{0+}$ 
in the charge channels $\gamma(Z) + p \to \pi^0 + p$ (left panels) 
and $\gamma(Z) + p \to \pi^+ + n$ (right panels) in units of 
($10^{-3}/m_\pi$) as function of $W$ in GeV. The upper panels 
show the real part, while the lower panels show the imaginary part 
of the multipole. The thick (thin) curves display results for 
$Q^2=0$ ($Q^2=0.5$ GeV$^2$), respectively. The dotted curves 
correspond to the electromagnetic result from MAID, the dashed 
curve represent the result for the weak NC multipole neglecting 
the strangeness contribution, whereas the solid curves show the 
full weak NC multipole including strangeness.
}
\label{fig1}
\end{figure}
\begin{figure}
\includegraphics[width=16cm]{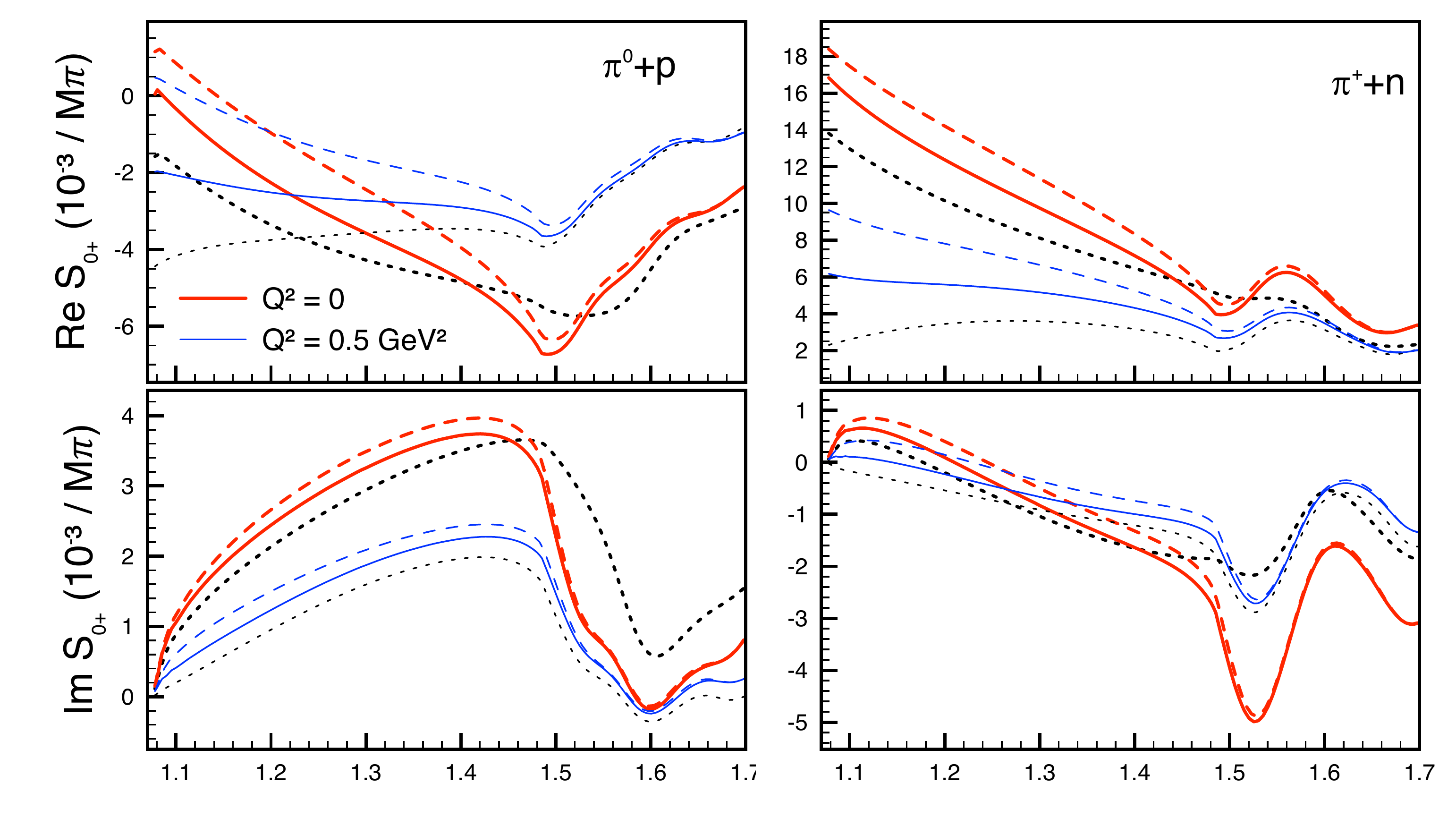}
\caption{Same as in Fig.\ \ref{fig1} for the multipole $S_{0+}$.}
\label{fig2}
\end{figure}
\begin{figure}
\includegraphics[width=16cm]{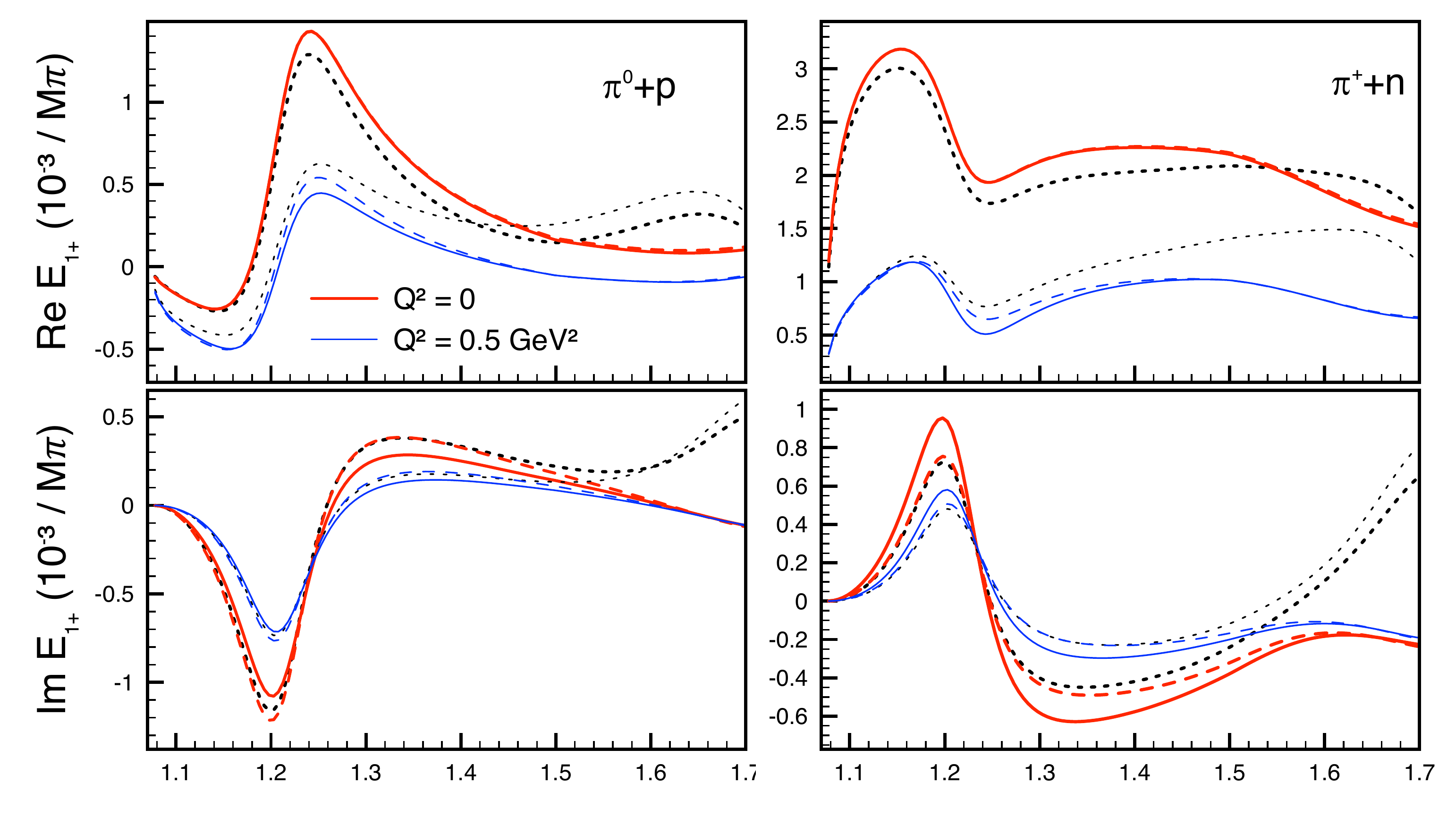}
\caption{Same as in Fig.\ \ref{fig1} for the multipole $E_{1+}$.}
\label{fig3}
\end{figure}
\begin{figure}
\includegraphics[width=16cm]{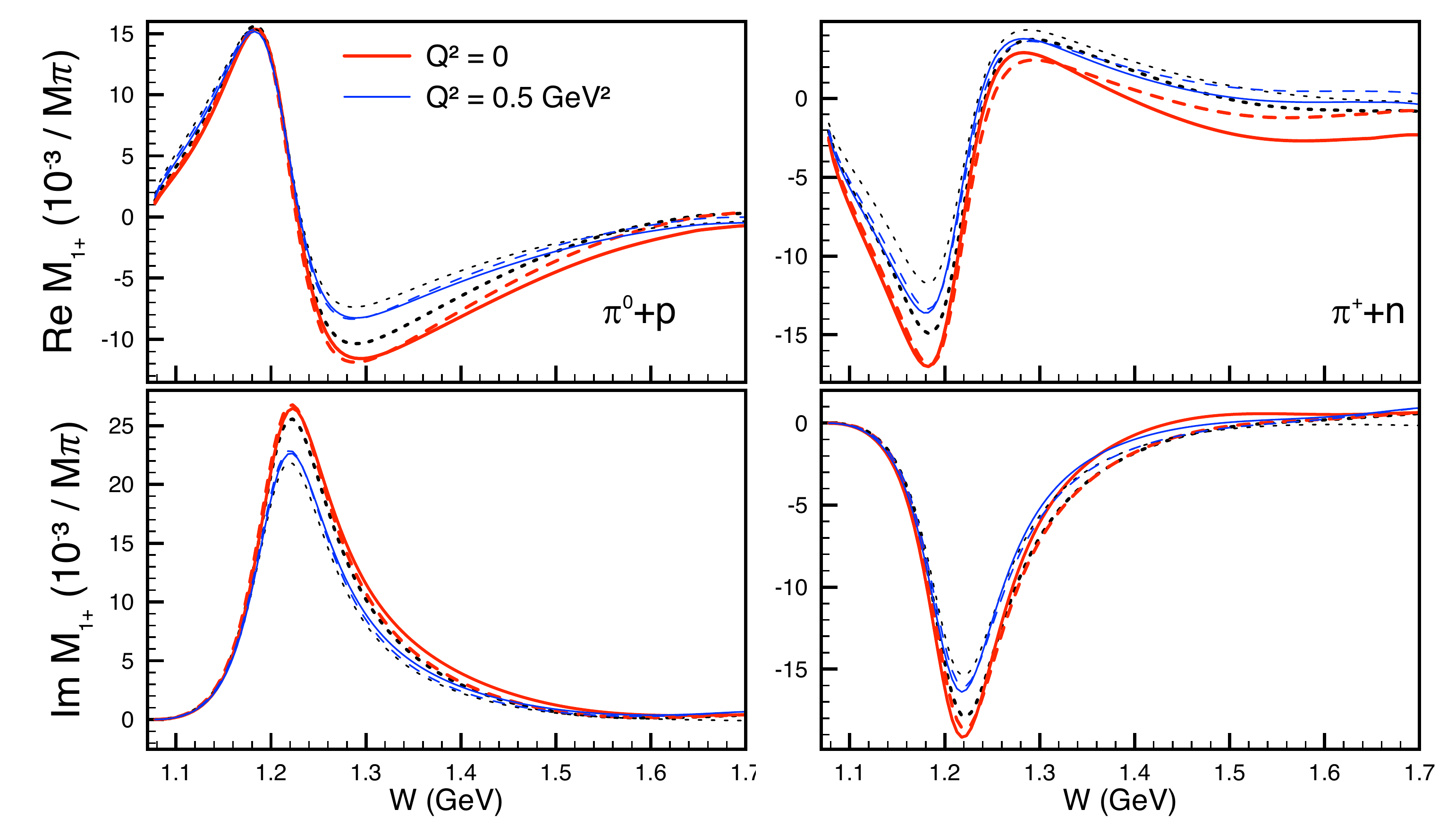}
\caption{Same as in Fig.\ \ref{fig1} for the multipole $M_{1+}$.}
\label{fig4}
\end{figure}
\begin{figure}
\includegraphics[width=16cm]{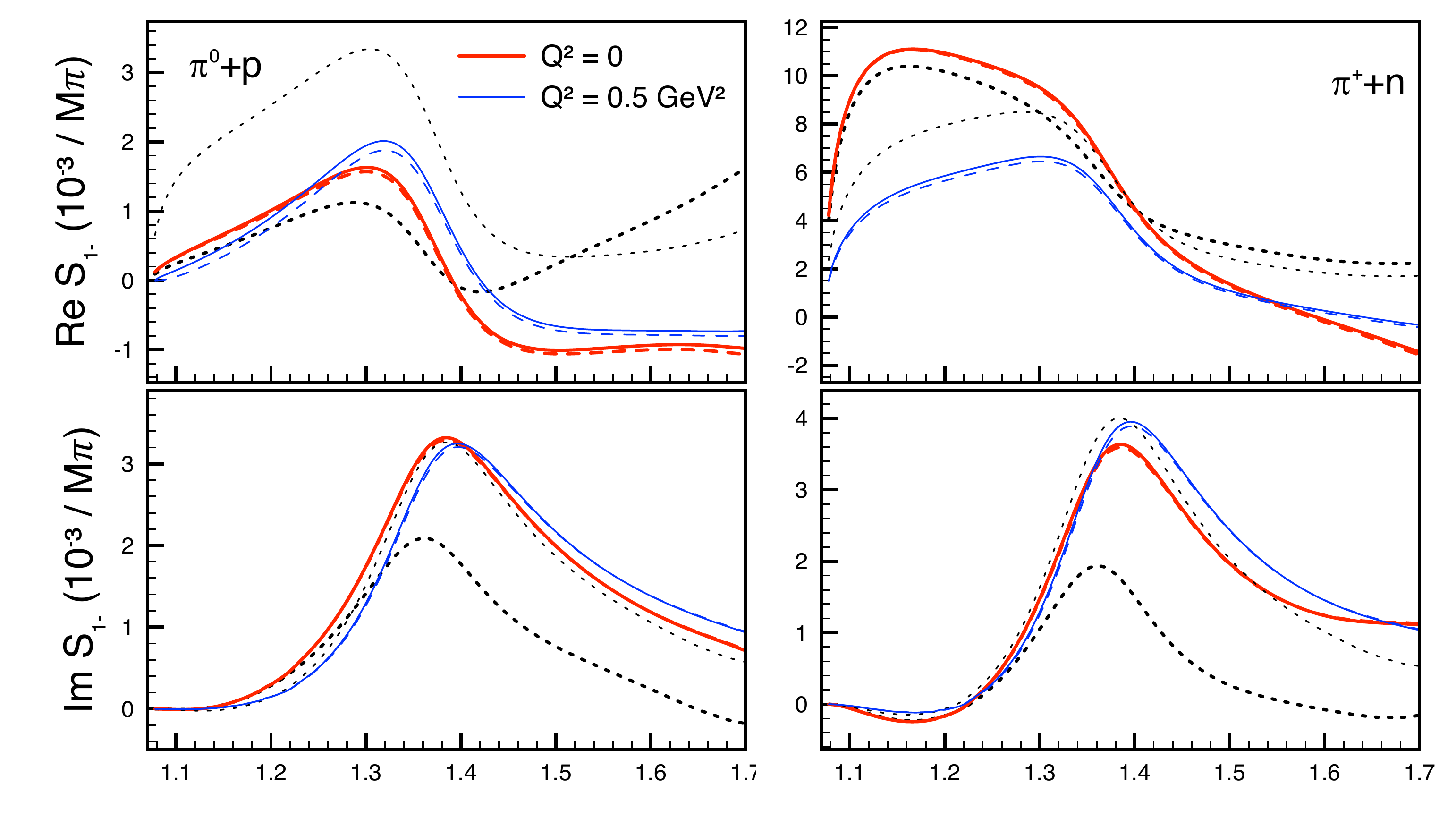}
\caption{Same as in Fig.\ \ref{fig1} for the multipole $S_{1-}$.}
\label{fig5}
\end{figure}
\begin{figure}
\includegraphics[width=16cm]{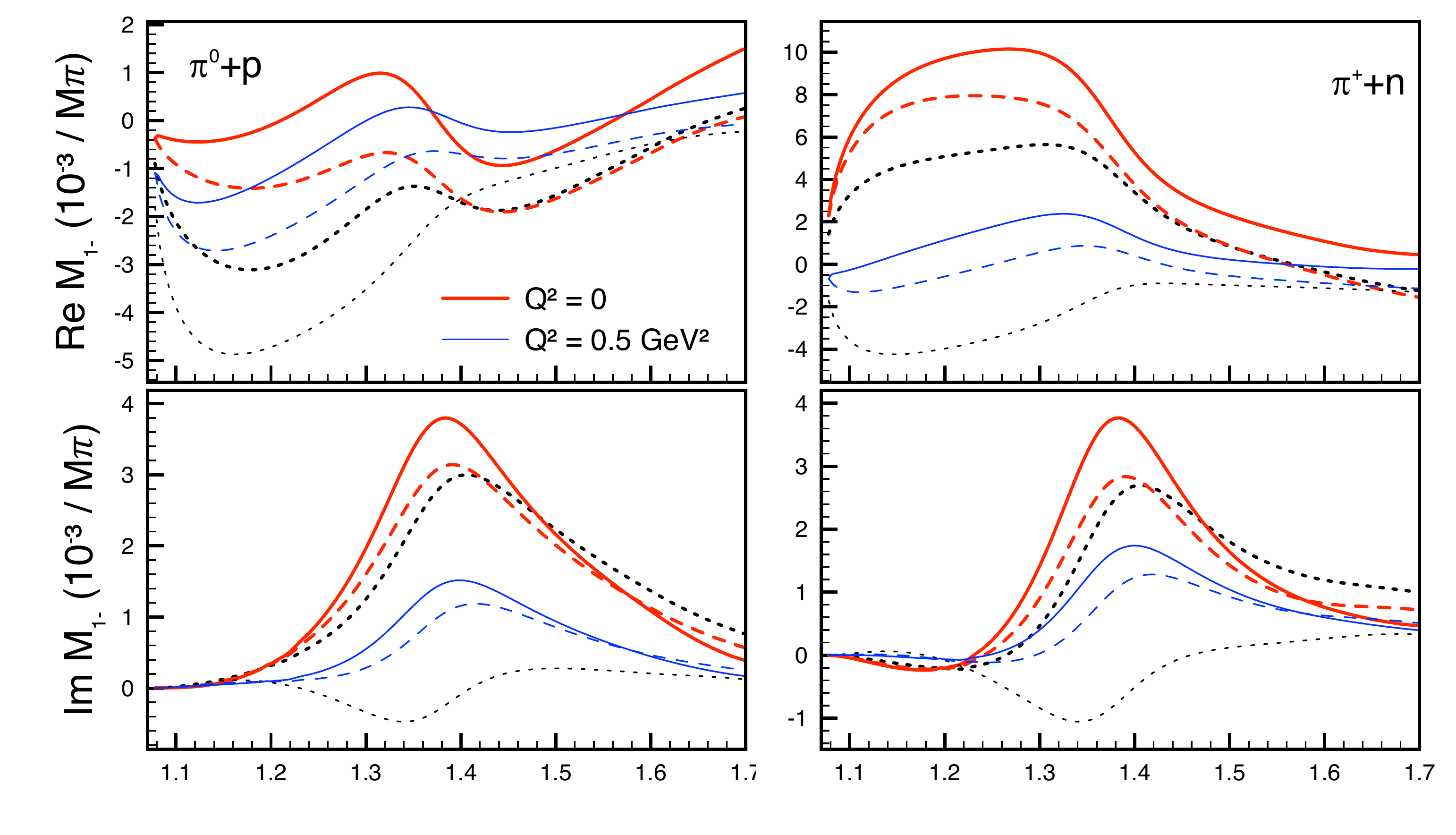}
\caption{Same as in Fig.\ \ref{fig1} for the multipole $M_{1-}$.}
\label{fig6}
\end{figure}
%


\section{Sensitivity of the PV asymmetry to $\mu_s$ at  
threshold and in the $\Delta$-region}
\label{sec:apv}

Weak pion production can be accessed, e.g., in parity-violating 
electron scattering (PVES). We will consider inclusive scattering 
here, i.e., we assume that only the electron in the final state 
is detected. In terms of the electromagnetic and $\gamma Z$ 
interference cross sections $\sigma_{T,L,A}^{\gamma,\gamma Z}$ 
the asymmetry for scattering of a beam of longitudinally 
polarized electrons off unpolarized protons reads
\beqn
A^{PV} &=& 
- \frac{G_FQ^2}{4\sqrt2 \pi\alpha_{em}} 
\frac{\sigma_T^{\gamma Z} + \epsilon\sigma_L^{\gamma Z} 
+ (1-4\sin^2\theta_W)\sqrt{1-\epsilon^2}\sigma_A^{\gamma Z}}%
{\sigma_T^{\gamma}+\epsilon\sigma_L^{\gamma}} \, ,\label{eq:apv}
\eeqn
with the usual polarization parameter $\epsilon$ ranging from 0 
for backward scattering to 1 for forward scattering,
\beqn
\epsilon = 
\left[1+2(\nu^2+Q^2)/Q^2 \tan^2(\theta/2)\right]^{-1} \, .
\eeqn
$\nu = E-E'$ is the energy of the virtual photon in the 
laboratory reference frame. The contribution from $\pi N$ final 
states, ${}^{\pi N}\sigma_{T, L, A}^{\gamma, \gamma Z}$, to 
these cross sections are expressed in terms of multipoles as
\beqn
{}^{\pi N}\sigma_T^{\gamma Z} &=& 
4\pi\frac{q}{k_\gamma^{CM}} \sum_{l}(l+1)^2 
\left[
\frac{l+2}{2}{\rm Re}(E_{l+}^{\gamma\,*}E_{l+}^Z 
+ M_{(l+1)-}^{\gamma\,*}M_{(l+1)-}^Z)
\right.
\nn\\
&& \qquad \qquad \qquad \qquad \quad
\left.
+ \frac{l}{2}{\rm Re}(M_{l+}^{\gamma\,*}M_{l+}^Z 
+ E_{(l+1)-}^{\gamma\,*}E_{(l+1)-}^Z)
\right] 
\, ,
\nn\\
{}^{\pi N}\sigma_L^{\gamma Z} &=& 
2\pi\frac{q}{k_\gamma^{CM}} \frac{Q^2}{\vec k\,^2}\sum_l(l+1)^3 
{\rm Re}(S_{l+}^{\gamma\,*}S_{l+}^Z 
+ S_{(l+1)-}^{\gamma\,*}S_{(l+1)-}^Z) 
\, , 
\nn\\
{}^{\pi N}\sigma_A^{\gamma Z} &=& 
4\pi\frac{q}{k_\gamma^{CM}} \sum_{l}(l+1)^2 
\left[
\frac{l+2}{2}{\rm Re}(E_{l+}^{\gamma\,*}{\cal{M}}_{l+}^Z 
+ M_{(l+1)-}^{\gamma\,*}{\cal{E}}_{(l+1)-}^Z)
\right.
\nn\\
&& \qquad \qquad \qquad \qquad \quad
\left.
+ 
\frac{l}{2}{\rm Re}(M_{l+}^{\gamma\,*}{\cal{E}}_{l+}^Z 
+ E_{(l+1)-}^{\gamma\,*}{\cal{M}}_{(l+1)-}^Z) 
\right] 
\, , 
\nn\\
{}^{\pi N}\sigma_T^{\gamma} &=& 
4\pi\frac{q}{k_\gamma^{CM}} \sum_{l}(l+1)^2 
\left[
\frac{l+2}{2}(|E_{l+}^\gamma|^2+|M_{(l+1)-}^\gamma|^2) 
+ 
\frac{l}{2}(|M_{l+}^\gamma|^2+|E_{(l+1)-}^\gamma|^2)
\right] 
\, , 
\nn\\
{}^{\pi N}\sigma_L^{\gamma} &=& 
2\pi\frac{q}{k_\gamma^{CM}} \frac{Q^2}{\vec k\,^2}\sum_l(l+1)^3 
\left[|S_{l+}^\gamma|^2+|S_{(l+1)-}^\gamma|^2\right] 
\, .
\eeqn
Here we have used $k_\gamma^{CM} = (W^2-M^2)/2W$ which stands 
for the real photon three-momentum in the center-of-mass frame. 
Expressions for ${}^{\pi N}\sigma_{T,L}^{\gamma}$ in terms of 
multipoles have been obtained, e.g., in Ref.\ \cite{Drechsel:1992pn}. 
In contrast to that reference, here the longitudinal cross section 
is written in terms of scalar, $S_{l\pm}$, rather than longitudinal 
multipoles, $L_{l\pm}$. The two sets of multipole moments are 
related by gauge invariance. Expressions for the interference 
cross sections ${}^{\pi N}\sigma_{T,L}^{\gamma Z}$ are a 
straightforward generalization of their electromagnetic 
counterparts, but to our knowledge have not been reported in 
the literature before, as is the case with the axial term 
${}^{\pi N}\sigma_{A}^{\gamma Z}$. Note that the longitudinal 
axial-vector multipoles do not enter 
${}^{\pi N}\sigma_{A}^{\gamma Z}$ which is hence purely transverse. 

As in Ref.\ \cite{Mukhopadhyay:1998mn}, the PV asymmetry defined 
in Eq.\ (\ref{eq:apv}) can be represented as a sum of three terms, 
\beqn
A^{PV} &=& 
- \frac{G_FQ^2}{4\sqrt2 \pi\alpha_{em}} 
\left[A_1+A_2+A_3\right] \, .
\eeqn
The first term, $A_1=2-4\sin^2\theta_W$, is model independent  
and results from isolating the isovector contribution in the 
numerator and denominator. The other two terms encode the 
isoscalar and strange ($A_2$) and the axial-vector ($A_3$) 
contributions. This form follows from the isospin decomposition 
of Eq.\ (\ref{eq:iso_Zpi}) but we do not explicitly rewrite 
Eq.\ (\ref{eq:apv}) here.


\begin{figure}
\includegraphics[width=16cm]{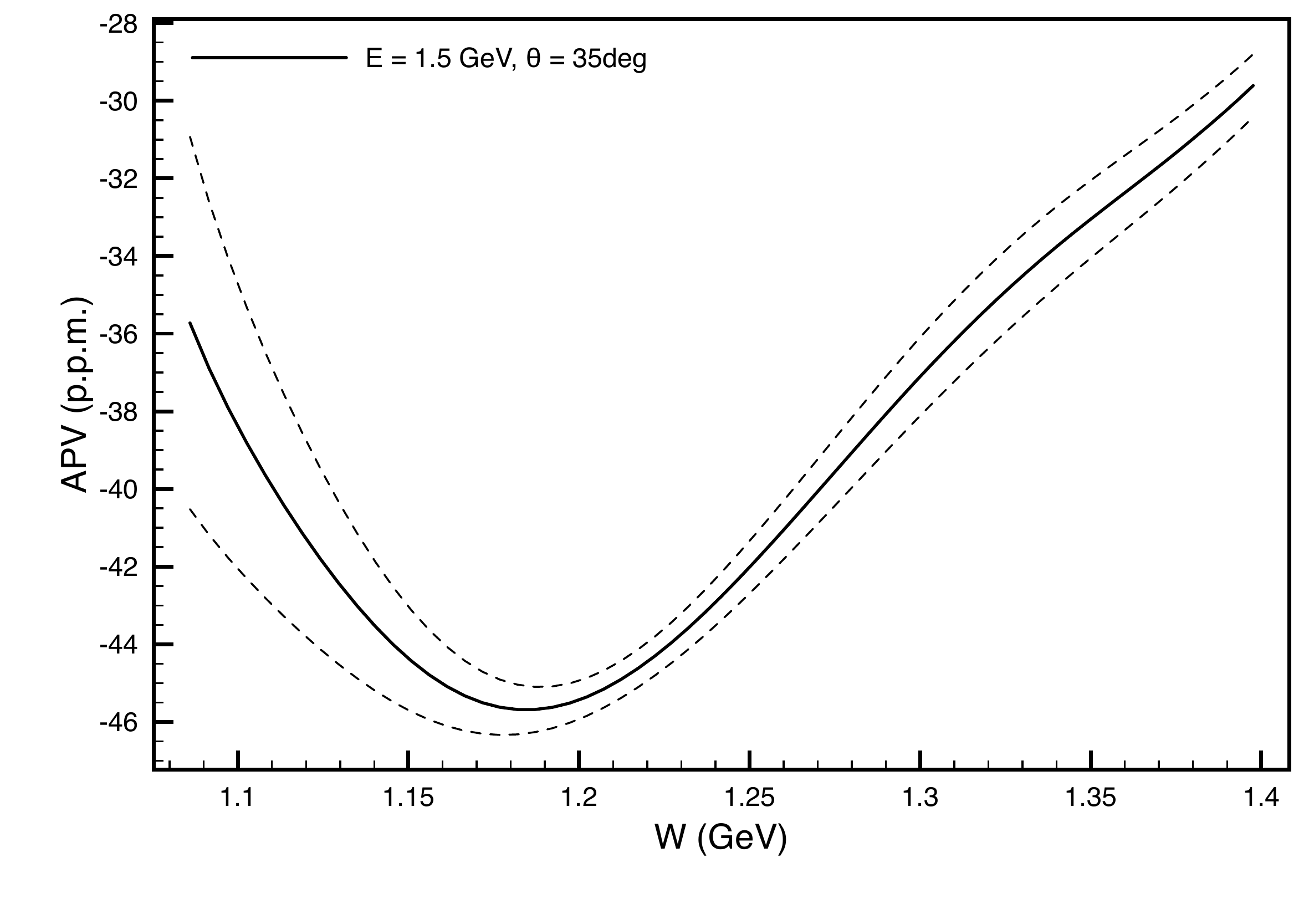}
\caption{PV asymmetry in the kinematics of A4@MAMI experiment with beam energy $E=1.5$ GeV and fixed electron scattering angle $\theta_e=35^\circ$. The various curves demonstrate the sensitivity to strange magnetic moment: the solid curve corresponds to the central value $\mu_s=0.38$, and the band between the upper and lower dashed curves indicates the uncertainty range $\delta\mu_s=\pm0.27$.}
\label{fig8}
\end{figure}

\begin{figure}
\includegraphics[width=16cm]{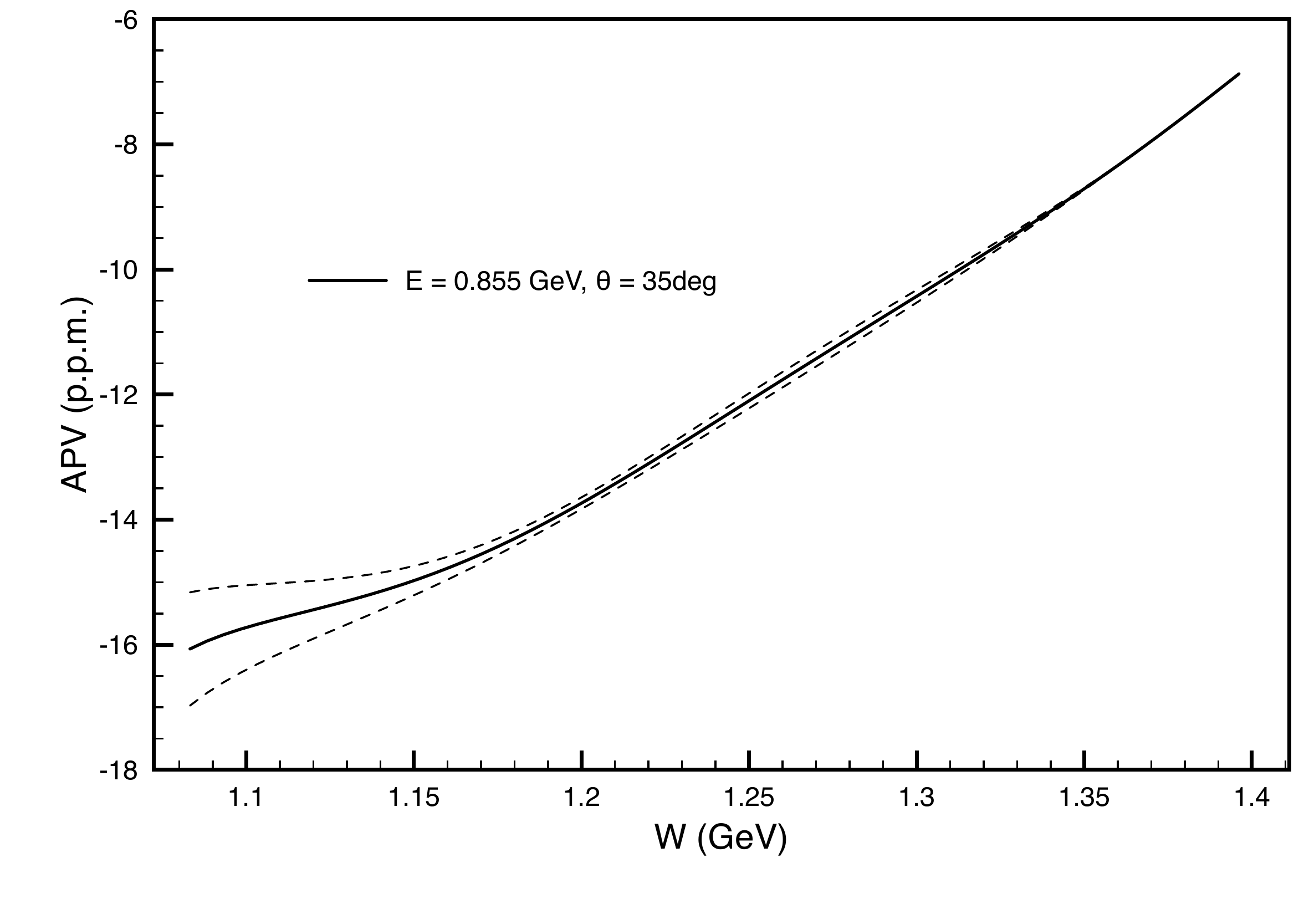}
\caption{Same as in Fig.\ \ref{fig8} for $E=0.855$ GeV.}
\label{fig9}
\end{figure}

We are interested in studying the sensitivity of the 
parity-violating asymmetry to strangeness contributions. 
To simplify the discussion and to keep the analysis 
uncontaminated, we start with results without taking the 
axial multipoles into account. In fact, their contribution 
is suppressed by the small weak charge of the electron, with 
$1 - 4 \sin^2\theta_W(0) \approx 0.048$. In addition, the factor 
$\sqrt{1 - \epsilon^2}$ leads to a further suppression for 
forward kinematics; however, this suppression is not there at 
backward angles. 

In Figs.\ \ref{fig8} and \ref{fig9}, we plot our results as 
a function $W$ in the range between the pion production threshold 
and the $\Delta$-resonance region assuming forward kinematics 
as measured in the A4 experiment at MAMI. We see that in this 
kinematic range there is considerable sensitivity to the 
strangeness contribution at the level of 10 -- 20\%, as 
indicated by the difference between the solid and dashed lines 
in Figs. \ref{fig8}, \ref{fig9}. This sensitivity is quite 
promising in view of the high precision of the experimental 
data which are at the level of 5\% \cite{A4resPV}. A targeted 
analysis of the available inelastic PVES data between threshold 
and the $\Delta$ resonance region will offer an alternative way 
to constrain strangeness form factors of the nucleon, virtually 
independent of the conventional measurements of elastic PVES.

Historically, the parity-violating asymmetry 
in electron scattering has been proposed to measure the axial 
$N \to \Delta$ transition \cite{Mukhopadhyay:1998mn}. Numerically, 
its contribution at the $\Delta$ resonance position in the 
backward kinematics of the G0 experiment is at the level of 
6\% \cite{Androic:2012doa}. Our estimates show that at the 
$\Delta$ position an extraction of the axial $N \to \Delta$ 
transition would still be possible since the uncertainty due 
to strangeness is below 3\%. The largest sensitivity to strange 
form factors is, not unexpectedly, observed between the threshold 
and the $\Delta$ resonance.


\section{An update of $\gamma Z$-box graph corrections to 
elastic PVES experiments}
\label{sec:gammaZ}

The $\gamma Z$-box graph corrections to elastic PVES have 
attracted significant attention recently. They can be evaluated 
in an approach based on dispersion relations from $\gamma Z$ 
interference structure functions for inelastic $ep$ scattering. 
Their vector part, ${\rm Re}\,\Box^V_{\gamma Z}$, is sensitive 
to low-energy form factor input and depends strongly on the 
beam energy. For the kinematics of the Qweak experiment the 
box graph corrections exceed the previously claimed theory 
uncertainty by a factor of 8. Currently, the size of 
${\rm Re}\,\Box^V_{\gamma Z}$ has settled at $5.4 \times 10^{-3}$ 
\cite{Gorchtein:2008px,Sibirtsev:2010zg,Rislow:2010vi, 
Gorchtein:2011mz,Hall:2013hta}, but the uncertainty 
estimates range from $0.4\times10^{-3}$ \cite{Hall:2013hta} 
to $2.0\times10^{-3}$ \cite{Gorchtein:2011mz}. 
In the framework of forward dispersion relations, the 
$\gamma Z$-box correction obeys a sum rule in terms of the 
inclusive interference structure functions $F_{1,2}^{\gamma Z}$, 
as function of the lab energy of the electron,
\beqn
{\rm Re}\,\Box^V_{\gamma Z}(E) &=& 
\frac{\alpha_{em}}{2\pi M^2E} 
\int\limits_{(M+m_\pi)^2}^\infty dW^2 
\int\limits_0^\infty \frac{dQ^2}{1+Q^2/M_Z^2}
\left\{\left[\frac{1}{2E}\ln\left|\frac{E+\omega}{E-\omega}\right| 
- \frac{1}{\omega}\right]F_1^{\gamma Z}(W^2,Q^2) 
\right. \nn \\
& & 
\left. 
+ \frac{M}{Q^2}\left[\frac{E}{\nu}\left(1-\frac{Q^2}{4E^2}\right) 
\ln\left|\frac{E+\omega}{E-\omega}\right|
+ \ln\left|1-\frac{E^2}{\omega^2}\right| 
- \frac{Q^2}{M\omega}\right]F_2^{\gamma Z}(W^2,Q^2) 
\right\} \, ,
\eeqn
where $\nu = (W^2 - M^2 + Q^2)/2M$ is the virtual photon energy 
in the laboratory frame, and $\omega = (\nu + \sqrt{\nu^2+Q^2})/2$.
The integral requires knowledge of the structure functions 
$F_{1,2}^{\gamma Z}$ in the whole kinematic range in the two 
variables $W$ and $Q^2$. For a meaningful evaluation of this 
integral, the $\pi N$ contribution to the inclusive structure 
functions developed in the previous section needs to be extended 
to $W>2$ GeV and $Q^2>2$ GeV$^2$, and contributions from 
higher-mass states need to be included. First, we deal with 
extending the $\pi N$ state contribution to higher energies 
using the Regge theory framework. Assuming that at $W=2$ GeV 
the background contribution (mainly the vector meson exchanges) 
dominate over the resonances in the cross section, we can use 
the Regge theory expectation $F_1(W\to\infty) \sim 
(W^2/1\,{\rm GeV}^2)^{\alpha_M(0)}$ and $F_2(W\to\infty) \sim 
(W^2/W_0^2)^{\alpha_M(0)-1}$. The meson Regge trajectory 
$\alpha_M(t) = \alpha_M^0 + \alpha'_M t$ is taken at $t=0$, and 
the relevant $\rho$-$\omega$ trajectories correspond to 
$\alpha_\rho^0 \approx \alpha_\omega^0 \approx 0.5$. 
Correspondingly, the simple {\it Ansatz}  
\beqn
F_1^{\pi N}(W\geq2\,{\rm GeV},Q^2) &=& 
F_1^{\pi N}(2\,{\rm GeV},Q^2) 
\times \left(\frac{W}{2\,{\rm GeV}}\right) \, , 
\nn\\
F_2^{\pi N}(W\geq2\,{\rm GeV},Q^2) &=& 
F_2^{\pi N}(2\,{\rm GeV},Q^2) 
\times \left(\frac{2\,{\rm GeV}}{W}\right)
\eeqn
was adopted. The $Q^2$-dependence beyond $Q^2 = 2$ GeV$^2$ is 
assumed to follow a dipole $\sim[1+Q^2/\Lambda_V^2]^{-2}$ with 
$\Lambda_V = 0.84$ GeV$^2$, reminiscent of nucleon form factors, 
\beqn
F_{1,2}^{\pi N}(W,Q^2\geq2\,{\rm GeV}^2) &=& 
F_{1,2}^{\pi N}(W,2\,{\rm GeV}^2) 
\times
\left(\frac{\Lambda_V^2+2\,{\rm GeV}^2}{\Lambda_V^2+Q^2}\right)^2 
\, .
\eeqn

This model is not very sophisticated; but since the integral 
is saturated to 99\% by the range $Q^2 < 2$ GeV$^2$ the details 
of the model for higher $Q^2$ are irrelevant at the level of 
the required precision. With the model specified above the 
numerical evaluation of the $\pi N$ state contribution leads 
to 
\beqn
{\rm Re}\,\Box^{V,\,\pi N}_{\gamma Z}
(E=1.165\,{\rm GeV}) 
&=& 
(2.10\pm0.05)\times10^{-3} \, ,
\eeqn
with the uncertainty coming from varying the strange form factors 
within their error bars and treating the part of the integral 
from $Q^2 > 2$ GeV$^2$ as an additional uncertainty, the two 
errors added in quadrature. 

The $\pi N$-continuum is only one part of the proton inelastic 
spectrum. Further contributions are due to higher resonance 
excitations in channels other than $\pi N$, such as $2\pi N$ 
and $\eta N$, included in the original fit by Christy and Bosted  
\cite{Christy:2007ve}. Their contributions can be evaluated by 
switching off the $\pi N$-decay channel for the resonances, 
\beqn
{\rm Re}\,\Box^{V,\,{\rm Res.\;no\;}\pi N}_{\gamma Z} 
(E=1.165\,{\rm GeV}) 
&=& 
(0.35\pm0.15)\times10^{-3} \, .
\eeqn
Further details can be found in Ref.\ \cite{Gorchtein:2011mz}. 
Finally, to include the non-resonant part of multi-particle 
intermediate states we adopt a background model resembling the 
one described in Ref.\ \cite{Gorchtein:2011mz} which starts at 
the $2\pi$-production threshold. Then we find 
\beqn
{\rm Re}\,\Box^{V,\,{\rm Back.\;no\;}\pi N}_{\gamma Z}
(E=1.165\,{\rm GeV}) 
&=& 
(3.23\pm1.41)\times10^{-3} \, .
\eeqn

Adding these three numbers we obtain a new prediction for the 
$\gamma Z_V$ dispersive correction,
\beqn
{\rm Re}\,\Box^{V}_{\gamma Z}
(E=1.165\,{\rm GeV}) 
&=& 
(5.58\pm1.41)\times10^{-3} \, .
\eeqn
Due to a more precise account of the lowest part of the proton 
excitation spectrum we observe that the uncertainty is reduced. 

Similarly, we find for the energy range coverd by the upcoming 
MESA/P2 experiment in Mainz,
\beqn
{\rm Re}\,\Box^{V,\,\pi N}_{\gamma Z} 
(E=0.155\,{\rm GeV}) 
&=& 
(0.60\pm0.02)\times10^{-3} \, , 
\nn\\
{\rm Re}\,\Box^{V,\,{\rm Res.\;no\;}\pi N}_{\gamma Z} 
(E=0.155\,{\rm GeV}) 
&=& 
(0.04\pm0.02)\times10^{-3} \, , 
\nn\\
{\rm Re}\,\Box^{V,\,{\rm Back.\;no\;}\pi N}_{\gamma Z} 
(E=0.155\,{\rm GeV}) 
&=& 
(0.43\pm0.18)\times10^{-3} \, ,
\eeqn
and for the total $\gamma Z$ box graph correction 
\beqn
{\rm Re}\,\Box^{V}_{\gamma Z} 
(E=0.155\,{\rm GeV}) 
&=& 
(1.07\pm0.18)\times10^{-3} \, .
\eeqn


\section{Conclusions}
\label{sec:conclusions}

To summarize, we studied the effect of the strangeness on the parity-violating pion electroproduction. We showed how the inclusion of the strange magnetic form factor modifies the Born contribution, obtained expressions for the vector multipoles for the subprocess $Z+p\to\pi+N$ and performed a unitarization procedure that follows the approach of Refs. \cite{Drechsel:1998hk,Drechsel:2007if}. At the moment we did not attempt to model the strangeness contributions to the $N\to N^*$ transition form factors, while no such contributions are present in the purely isovector $N\to\Delta$ transitions. This effectively limits the applicability of our model to energies between pion production threshold and the Roper resonance. We applied this model to the PV asymmetry in inclusive linearly polarized electron scattering off an unpolarized proton  target, and demonstrated that at higher energies and moderate $Q^2\sim0.6$ GeV$^2$, as in the kinematics of the A4@MAMI experiment, the sensitivity to the value of the strange magnetic moment is at the level 20\% relative to the asymmetry size of $\sim40$ p.p.m. At lower electron energy and correspondingly lower $Q^2$ this sensitivity decreases to $\sim10$\%. In view of the intrinsic statistical uncertainty of the asymmetry data at the level of 5-10\%, this sensitivity offers a quite promising new way to address strangeness with inelastic, rather than elastic PVES. The main advantage lies in much larger asymmetries than in the elastic case. Elastic and inelastic measurements  even with the same apparatus will have different systematics, making the extraction of the strange magnetic form factor from the data below and above the pion production threshold practically independent. The model for pion production with a weak probe allowed us to calculate the contribution of the $\pi N$-state to the inclusive interference structure functions $F_{1,2}^{\gamma Z}$ that enter the calculation of the dispersion $\gamma Z$-correction to the weak charge. This correction has to be taken into account for the extraction of the weak mixing angle from the elastic PVES data within the Q-Weak and P2 experiments. A more careful modeling of the lower part of the nucleon excitation spectrum done here allowed to shift the uncertainties to higher energies, leading a reduction of the uncertainty of the dispersive calculation of the energy-dependent correction $\Box^V_{\gamma Z}(E)$. In the Q-Weak kinematics $E=1.165$ GeV the new uncertainty estimate is $\delta\Box^V_{\gamma Z}(1.165{\rm GeV})=0.00141$, about 2\% of the Standard Model expectation for the proton's weak charge. For the P2 kinematics, $E=0.155$ GeV, the new uncertainty estimate is an order of magnitude smaller. 

\begin{acknowledgments}
M.G. and H.S. acknowledge useful discussions with F. Maas, K. Kumar, L. Capozza and M.J. Ramsey-Musolf, and the support by the Deutsche Forschungsgemeinshaft through the Collaborative Research Center ÒThe Low-Energy Frontier of the Standard ModelÓ CRC 1044. X.Z. acknowledges support from the US Department of Energy under grant DE-FG02-97ER-41014, and from Fermi National Accelerator Laboratory under intensity frontier fellowship.
\end{acknowledgments}

\begin{appendix}

\section{Vector Born multipoles with magnetic strangeness}

We list here for completeness the results for the multipoles 
for vector current pion production including strange magnetic 
form factors. We follow Ref.\ \cite{Drechsel:1998hk} in using 
a combination of pseudo-scalar and pseudo-vector $\pi NN$ 
couplings, see Eq.\ (\ref{eq:piNNmaid}). Note that the strange 
magnetism has no isovector component, therefore it only 
contributes to the $I=0$ isospin component of the multipoles.
We denote the orbital angular momentum of the $\pi N$ state by 
$l$, the strangeness magnetic form factor $F_2^s(Q^2)$, and we 
use the kinematic variables introduced previously in Section 
\ref{sec:kinematics}. The strange contributions to the vector 
multipoles then read
\beqn
E^{0,\,s}_{l+} 
&=& 
\frac{egNF_2^s(Q^2)}{16\pi W(l+1)}
\left[2\delta_{l0}
\left(\frac{1}{W+M}+\frac{W-M}{M^2}
\frac{\Lambda_m^2}{\Lambda_m^2+q^2}\right)
- (W-M)T_{l+}
\right.
\nn\\
&& 
\left.
- \frac{(W+M)}{M(E_1+M)}lR_l 
- \frac{(W-M)q}{Mk(E_2+M)}(l+1)R_{l+1}
\right] 
\, , 
\\
E^{0,\,s}_{l-} 
&=& 
\frac{egNF_2^s(Q^2)}{16\pi Wl}
\left[
- (W-M)T_{l-} 
+ \frac{(W+M)}{M(E_1+M)}(l+1)R_l 
+ \frac{(W-M)q}{Mk(E_2+M)}lR_{l-1}
\right], 
\\
M^{0,\,s}_{l+} 
&=& 
\frac{egNF_2^s(Q^2)}{16\pi Wl}
\left[
- (W-M)T_{l+} 
+ \frac{(W+M)}{M(E_1+M)}R_l
\right] \, , 
\\
M^{0,\,s}_{l-} 
&=& 
\frac{egNF_2^s(Q^2)}{16\pi W(l+1)}
\left[ 
- \frac{2qk\delta_{l1}}{N^2} 
\left(\frac{1}{W-M} 
+ \frac{W+M}{M^2}\frac{\Lambda_m^2}{\Lambda_m^2+q^2}
\right) 
\right.
\nn\\
&& \left.
+ 
(W-M)T_{l+} 
- \frac{(W+M)}{M(E_1+M)}R_l
\right] \, , 
\\
S^{0,\,s}_{l+} 
&=& 
\frac{egNF_2^s(Q^2)}{16\pi W(l+1)}
\left\{ 
\frac{k\delta_{l0}}{M^2}
\left( 
- \frac{M}{W+M}\left(1+\frac{W+M}{E_1+M}\right) 
+ 2\frac{\Lambda_m^2}{\Lambda_m^2+q^2}\right) 
\right. \\
&& 
\left.
- (-1)^{l} \left[ 
(q^0-\frac{k^0}{2}) 
\left(\frac{Q_l(\bar E_2)}{q(E_1+M)} 
+ \frac{Q_{l+1}(\bar E_2)}{k(E_2+M)} 
\right)
\right.\right.
\nn\\
&& \left.\left.
- \frac{Wk^0+Q^2-m_\pi^2}{2M} 
\left(\frac{Q_l(\bar E_2)}{q(E_1+M)} 
- \frac{Q_{l+1}(\bar E_2)}{k(E_2+M)}\right)
\right]
\right\} 
\, , 
\nn \\
S^{0,\,s}_{l-} &=& 
\frac{egNF_2^s(Q^2)}{16\pi Wl}
\left\{ 
- \frac{qk^2\delta_{l1}}{M^2N^2} 
\left(\frac{M}{W-M}\left(1+\frac{W-M}{E_1-M}\right)
+ 2\frac{\Lambda_m^2}{\Lambda_m^2+q^2}
\right) 
\right. 
\\
&& 
\left.
- (-1)^{l} \left[ 
(q^0-\frac{k^0}{2}) 
\left(\frac{Q_l(\bar E_2)}{q(E_1+M)} 
+ \frac{Q_{l+1}(\bar E_2)}{k(E_2+M)} 
\right)
\right.\right.
\nn\\
&& \left.\left.
- \frac{Wk^0+Q^2-m_\pi^2}{2M} 
\left(\frac{Q_l(\bar E_2)}{q(E_1+M)} 
- \frac{Q_{l+1}(\bar E_2)}{k(E_2+M)}
\right)
\right]
\right\} \, .
\nn
\eeqn
We have also used the abbreviations 
\beqn
N &=& \sqrt{(E_1+M)(E_2+M)} \, , 
\nn\\
Q_l(y) &=& \frac{1}{2}\int_{-1}^1dx\frac{P_l(x)}{y-x} \, , 
\nn\\
R_l &=& \frac{(-1)^l}{2l+1} 
\left[Q_{l+1}(\bar E_2)-Q_{l-1}(\bar E_2)\right] \, , 
\nn\\
T_{l\pm} &=& 
(-1)^l\left[\frac{Q_{l}(\bar E_2)}{kq} 
- \frac{W+M}{(E_1+M)(E_2+M)(W-M)}Q_{l\pm1}(\bar E_2)\right] \, .
\eeqn
$P_l(x)$ are the Legendre polynomials, $Q_l(x)$ the 
associate Legendre polynomials, and $\bar E_2 = 
(2k^0E_2+Q^2)/2kq$.


\end{appendix}


\end{document}